\documentclass[twocolumn,showpacs,aps]{revtex4}
\usepackage{epsf}

\def\div{{\mathrm{div }  }}
\def\hb{\hbar}

\def\bO{{\bf \Omega}}
\def\tO{{\bf {\tilde \Omega}}}
\def\ep{\varepsilon}
\def\lep{{|\mathrm{log }\ \ep|}}
\def\ln{{\mathrm{log }}}

\def\dist{{\rm dist}}

\def\nab{\nabla}
\def\np{\nab^{\perp}}

\def\O{\Omega}

\def\a0{\alpha_0}
\def\a{\alpha}
\def\too{\tilde\Omega}
\def\la{\lambda}
\def\be{\begin{equation}}
\def\ee{\end{equation}}
\def\beq{\begin{equation}}
\def\eeq{\end{equation}}

\def\ba{{\bf A}}
\def\br{{\bf r}}

\def\cd{{\cal D}}
\def\ce{{\cal E}}

\begin{document}

\title{Vortices in a rotating
Bose-Einstein condensate:\\ 
critical velocities and energy diagrams in 
 the Thomas-Fermi regime}

\author{Amandine Aftalion}
\email{aftalion@ann.jussieu.fr}
\affiliation{Laboratoire d'analyse num\'erique,
B.C.187,  Universit\'e Paris 6, 175 rue du Chevaleret, 75013 Paris,
France.}
\author{Qiang Du}\thanks{Also at the Department of Mathematics, Hong Kong
University of Science and Technology,  Clear Water Bay, Hong Kong}
\email{qdu@iastate.edu}
\affiliation{Department of Mathematics, Iowa State University,
Ames, IA 50011, USA.}
\date{\today}


\begin{abstract}
For a Bose-Einstein condensate placed in a rotating trap and confined
in the $z$ axis, we set a framework of study for the Gross-Pitaevskii
energy in the Thomas Fermi regime. We 
investigate an asymptotic development of the energy, the critical
velocities of nucleation of vortices with respect to a small
parameter $\ep$ and the location of vortices. The limit $\ep$ going to
zero corresponds to the Thomas Fermi regime.
The non-dimensionalized energy is  similar to the Ginzburg-Landau 
energy for superconductors in the high-kappa high-field limit 
and our estimates rely on techniques developed for this latter problem. 
We also take the advantage of this similarity to develop a numerical 
algorithm for computing the Bose-Einstein vortices. Numerical results 
and energy diagrams are presented.
\end{abstract}

\maketitle

\section{Introduction}

Since the first experimental achievement of Bose-Einstein condensates in
atomic gases in 1995, many properties of these systems have been studied
experimentally and theoretically, and particularly the existence of
vortices \cite{BuR,CD,DGPS,FCS,F,FS,MCWD,SF}. 
A way to create vortices consists in rotating the trap confining the
atoms: for sufficiently large velocities, it becomes energetically
favorable to have vortices in the system. 
Theoretical studies of this type of
experiments have often been made
in the framework
of the nonlinear Schr\"odinger equation or Gross-Pitaevskii equation,
well known for superfluids, but which provides a very good description
of Bose-Einstein condensates: it is assumed that the $N$ particles of
the gas are condensed in the same state for which the wave function
$\phi$ minimizes the Gross-Pitaevskii energy. 
By introducing a rotating frame at the angular
velocity $\tO = \too{\bf e}_z$,  the trapping potential becomes time
independent, and the wave function $\phi$ minimizes the energy
\begin{eqnarray}
\label{BE}
{\cal E}_{3D}(\phi) &=& \int {\hb^2 \over{2m}} |\nabla \phi|^2 +{m\over 2}
 \sum_{\alpha} \omega^2_\alpha  r^2_\alpha |\phi|^2\nonumber\\
&& +{N\over 2} g_{3D}|\phi|^4-{{\hb {\tO}}}\cdot
(i\phi, \nabla \phi\times {\bf x}) , 
\end{eqnarray}
under the constraint $\int |\phi|^2 =1$. Here,
for any complex quantities $u$ and $v$ and their
complex conjugates $\bar{u}$ and $\bar{v}$, 
 $(u,v)=(u\bar{v}+\bar{u}v)/2$.
The terms in the energy correspond to the kinetic energy, the trapping
potential energy, the interaction energy and the inertial due to the
change of frame.   

When the atoms are strongly confined in the $z-$axis, the situation can
be simplified to a two dimensional problem where the wave function
$\psi$ depends only on ${\bf x}=(x,y)$ and it minimizes
\begin{eqnarray}
\label{BE2D}
{\cal E}_{2D}(\psi) &=& \int {\hb^2 \over{2m}} |\nabla \psi|^2 +{m\over 2}
 \sum_{\alpha=x,y} \omega^2_\alpha  r^2_\alpha |\psi|^2\nonumber\\
&& +{N\over 2} g|\psi|^4-{{\hb
{\tO}}}\cdot (i\psi, \nabla \psi\times {\bf x}) , 
\end{eqnarray}
where $g=g_{3D}(m\omega_z/2\pi \hb)$. 
The  constraint $\int |\psi|^2 =1$ is also imposed. Our study was at first
motivated by the work of Castin and Dum \cite{CD} who have studied the
equilibrium configurations by looking for the minimizers in a reduced
class of functions for the 2D case and have done numerical computations
in 2 and 3D. Their analysis is 
in the Thomas-Fermi regime, where the mean interaction energy per
particle is larger than $\hb \omega_{x,y}$. 

Our aim is to provide a mathematical
framework for a rigorous study of the energy ${\cal
E}_{2D}$ and its minimizers in the Thomas Fermi limit. 
We first observe that
this energy has a striking similarity with the high-kappa, high-field
limit  of the Ginzburg-Landau free energy used in the
modeling of superconductors. Thus, we expect
the energy $\ce_{2D}$ will develop similar behavior 
as those for
the Ginzburg-Landau energy 
studied in \cite{ASS,S1,S}. In particular, the results obtained
in the context of Ginzburg-Landau energy may be applied to 
${\cal E}_{2D}$, in the Thomas Fermi regime, to yield an asymptotic
development of the energy as well as the critical
velocities for the nucleation of vortices and the location of these
vortices. 
Due to the close resemblance, we will not be concerned with the detailed
derivations in this paper, but rather focus on the conclusions one can
draw from the asymptotic developments.  To our knowledge, some of our
estimates to be presented later have not been given in the literature
previously. Let us point out that our method 
to compute the energy is also new in this context and very different
from the ones in, for instance, \cite{CD,FCS,SF}.

We define the characteristic length  $d=(\hb / m\omega_x)^{1/2}$ and set
$$\ep^2= {{\hb ^2}\over {2Ngm}}.$$ In the Thomas-Fermi approximation, $\ep$ is
small, which will be our asymptotic regime. 
We re-scale the distance by $R=d/\sqrt \ep$ and
define $u({\bf r})=R \psi({\bf x})$ where ${\bf x}=R{\bf r}$. Assume
that $\omega=\omega_x$ and $\omega_y=\lambda\omega$ with $0\leq \lambda
\leq 1$ and we set $\O =\too /\ep\omega$.
Since the trapping potential is stronger than the inertial potential, we
have $\O <1 / \ep$. The energy can be rewritten as:
\begin{eqnarray}
\label{be2d}
{E}_{2D}(u) &=&\int {1\over 2} |\nabla u|^2+{1\over
2\ep^2}(x^2+\lambda^2y^2)|u|^2+{1\over 4\ep^2}|u|^4\nonumber\\
&&+ \bO \cdot (iu,\nabla
u\times {\bf r}) \; .
\end{eqnarray}
Due to the constraint $\int |u|^2 =1$,  we can add to ${ E}_{2D}$
any multiple of 
$\int
|u|^2$ so that it is equivalent to minimize 
$$\int |\nabla u |^2+2 \bO \cdot (iu,\nabla u\times {\bf r})+{1\over
2\ep^2}|u|^4-{1\over \ep^2}a({\bf r})|u|^2$$ 
where $a({\bf r})=\a -(x^2+\lambda^2y^2)$ for some constant $\a$ 
to be determined.  Let $\cd$ be the ellipse
$\{a >0\}=\{x^2+\lambda^2 y^2<\a\}$. We impose the following constraint
on  $\a$:
\beq\label{cons}\int_{\cd}a({\bf r})=1.\eeq
Indeed, as $\ep$ tends to 0, the minimizer will satisfy that $|u|^2$ will
be close to $a$ so that the constraint will be satisfied automatically by $u$ if we impose (\ref{cons}). Equation
 (\ref{cons})
leads to $\a^2=2\la / \pi$. If $\lambda=1$, that is $\omega_x=\omega_y$,
then $\cd$ is a disc of radius $R_0$ with $R_0^4=2/\pi$. 

To study the problem analytically, it is reasonable to minimize
the energy over the domain $\cd$ with zero boundary data for
$u$. Indeed, when $a\leq 0$,  the energy is convex so that the
minimizer $u$ goes to zero exponentially at infinity (see  the numerical
observation in \cite{CD} and the analysis on the behavior near the
boundary of $\cd$  as well as the decay at infinity of the order
parameter in \cite{DPS,FF}). 
Denote $H^m(\cd)$ the space of square integrable functions
defined on the domain $\cd$  that 
have square integrable derivatives up to the order $m$, and
 $H^1_0(\cd)$ the space of functions in $H^1(\cd)$ satisfying
the zero boundary condition. Denote the norm
$\left(\int_{\cd} |v|^2 \right)^{1/2}$ by $\|v\|$ for any square
integrable function $v$, 
 we then  consider  the problem
$$\min E_\ep(u)\; \;\hbox{subject to }\;
u\in H_0^1(\cd),\; \|u\| =1\; 
\eqno{(P)}$$
where
\begin{eqnarray}
\label{be}
  E_\ep (u)&=&\int_\cd |\nabla u |^2+ 2 \bO \cdot (iu,\nabla u\times {\bf
r})\nonumber\\
&& +{1\over 2\ep^2}(a({\bf r})-|u|^2)^2 .
\end{eqnarray} 
Note that critical point $u$ of $E_\ep$ is a solution of
\beq\label{eqbe}
-\Delta u-2i(\bO\times {\bf r}).\nabla u={1\over \ep^2}u(a-|u|^2)+\mu_\ep u 
 \hbox{ in }  \cd,
\eeq
with $u=0$ on $\partial \cd$ and $\mu_\ep$ is the Lagrange multiplier. The specific choice of $\a$ in (\ref{cons}) will imply that the term $\mu_\ep u$ is negligible in front of $au/\ep^2$.

We want to study the behavior of $\min  E_\ep(u)$ as $\ep$ goes
to 0.  In Section 2, we compute an asymptotic development of the
energy and in Section 3, the critical velocities of nucleation of
vortices and the location of the vortices.  
In Section 4, we study the evolution in imaginary time and construct
some numerical algorithms. 
In Section 5, we present some computational
results and the energy diagrams.

\section{Asymptotic development of the energy}

To study the behavior of the minimizer of the energy when $\ep$
goes to zero,  
we observe that the form of the energy (\ref{be}) is close to the 
Ginzburg-Landau functional studied in \cite{ASS}, \cite{S1} 
where the magnetic field has been replaced by a rotating term and
similar to \cite{S} except for the trapping potential and the
minimization over constraint. 
The main idea
 is to decouple the energy
into three terms: a part coming from the solution without vortices, a vortex
contribution and a term due to rotation. The estimate on the vortex
contribution was developed in \cite{BBH,BR,S1,S}. 

\subsection{The solution without vortices}
Firstly, we are interested in solutions without vortices, that is $u$
has no zero in the interior of $\cd$. 
Thus we consider functions of the form $\eta=fe^{iS}$,
where $\eta$ is in $H_0^1(\cd)$ and $f$ is real and has no zero in
the interior of $\cd$. We consider first minimizing $E_\ep$ over
such functions without imposing the
constraint that the $L^2$ norm is 1,
that is, $f$ and $S$ minimize
\begin{eqnarray}\label{fSener}
{\cal E}_\ep(f,S)&=&\int_{\cd} |\nabla f |^2+{1\over
2\ep^2}(a-f^2)^2\nonumber\\
&&+\int f^2|\nabla S-\bO\times {\bf r}|^2-f^2\O^2r^2. 
\end{eqnarray}
We have $f=0$ on $\partial \cd$ and 
\beq\label{f}
-\Delta f+f\nabla S(\nabla S-2\bO\times {\bf r})={1\over \ep^2}f(a-f^2)
\hbox{ in } \cd , 
\eeq
\beq\label{S}
\div (f^2(\nabla S-\bO\times {\bf r}))=0.\eeq
Equation (\ref{S}) implies that there exists $\xi$ in $H^2(\cd)\cap
H^1_0(\cd)$ such that 
\beq\label{xinp}
f^2(\nabla S-\bO\times {\bf r})=\O \np\xi, \eeq
where $\np\xi=(-\partial_y\xi, \partial_x \xi)$. So $\xi$ is the
unique solution of 
\beq\label{xieq}
\div ( {1\over f^2} \nabla\xi)=0 \hbox{ in } \cd, \quad \xi=0 \hbox{ on
} \partial \cd.\eeq 
Note that equation (\ref{xinp}) is the equation for the velocity, but we
prefer to write it as an orthogonal gradient for our later purposes of
integration by part. 
In the special case where $\cd$ is a disc, the minimum of
(\ref{fSener}) is reached for $\nabla S=0$ but this is not the case if
$\cd$ is an ellipse and there is a non trivial solution of (\ref{S}). 

\paragraph{The case of the disc}
Assume that $\la=1$ so that $\cd$ is a disc and
$a (r)=R_0^2-r^2 $. As discussed earlier, $\nabla S=0$ 
 in this case so that the energy becomes
\beq\label{enerreal}
{\cal E}_\ep(f)=
\int_{\cd} |\nabla f |^2+{1\over 2\ep^2}(a({ r})-f^2)^2.
\eeq 
Let $\eta_\ep$ be the minimizer of
(\ref{enerreal}) over real valued functions in $H^1_0(\cd)$.
Then, $\eta_\ep$ has no vortex, is independent of $\O$ and  satisfies
$$\Delta \eta_\ep ={1\over \ep^2}\eta_\ep ( \eta_\ep^2-a)\ \hbox{in}\
\cd,\quad \eta_\ep=0\ \hbox{on}\ \partial \cd.$$ 
When $\ep$ tends to 0, $\eta_\ep^2$ is close to $a$ except on a boundary
layer of size $\ep^{2/3}$ close to $\partial \cd$.
More precisely, using sub and super solutions, one can verify that 
$$\sqrt {a(r)}\tanh({\delta (\sqrt {a(r)})^3\lep})\leq \eta_\ep\leq
\sqrt {a(r)}$$
for $|R_0^2-r^2|\geq C\ep^{1/3}$.  
In fact, one can construct a sub solution
of the type above in any region $|R_0^2-r^2|\geq C\ep^{\beta/3}$ with
$\beta<2$. Then the value of $\delta$ is less than $c(2-\beta)$. 

The boundary layer can be analyzed using the change of variable
$x=(R_0-r)/\ep^{2/3}$ and $v_\ep(x)=\eta_\ep (r) /\ep^{2/3}$. Then
$v_\ep$ satisfies the Painlev\'e equation 
$$v''=v(v^2-2R_0x), \;\; v(0)=0,\;\; v(x)\simeq 2R_0x \;\; \hbox{for}
\ x \ \hbox{large}.$$ 
The boundary behavior has already been studied in 
\cite{FF} and \cite{DPS}, but using matched asymptotics. 

The energy of $\eta_\ep$  can also be estimated by a test function
equal to $\sqrt a$ except on the boundary layer to get 
\beq\label{upperbdener}
E_\ep(\eta_\ep)\leq {{2\pi}\over 3}\lep (1+o(1)).
\eeq

\paragraph{The case of an ellipse}
As discussed before, the minimum $\eta_\ep=
f_\ep e^{iS_\ep}$ of (\ref{fSener}) has a non-trivial phase.  
$f_\ep^2$ tends to $a$ in every compact subset of $\cd$ and
the function $\xi_\ep$ given by (\ref{xinp}) or (\ref{xieq})  tends to
the unique solution $\xi$ of  
\beq\label{xilim}
\div({1\over a}\nabla \xi)=0 \hbox{ in } \cd, \quad \xi=0 \hbox{ on }
\partial \cd.\eeq 
One can easily get that $\xi(x,y)={-a^2(x,y)/(2+2\la^2)}$.
Using (\ref{xinp}), we can define
 $S_0$, the limit of $S_\ep$,
 to be the solution of  $a(\nabla S_0-\bO\times{\bf r})=\O\np \xi$
with zero value at the origin. We have $S_0=C\O xy$ with
$C=(\la^2-1)/(\la^2+1)$. We see that $S_0$ cancels when $\la=1$ that is in
the case of the disc.   This computation is consistent with the one in
\cite{SF}, though it is derived in a different way. 

\subsection{Decoupling the energy}
Let $\eta_\ep=f_\ep e^{iS_\ep}$ be the vortex free 
minimizer of $E_\ep$ discussed previously without imposing
the constraint on the norm of $u$. Let
$u_\ep$ be a minimizer of $E_\ep$ under the constraint $\int_{\cd} |u|^2=1$
and  let $v_\ep=u_\ep/\eta_\ep$. Since $\eta_\ep$  satisfies the Gross
Pitaevskii equation (\ref{f})-(\ref{S}), we have 
\begin{eqnarray}
&&\int_{\cd} (|v|^2-1) (-{1\over 2}\Delta f_\ep^2-{1\over \ep^2}f_\ep^2
(a - f_\ep^2)+|\nabla f_\ep e^{iS_\ep}  |^2\nonumber\\
&&\quad\quad-2f_\ep ^2(\nab S_\ep\cdot
\bO\times {\bf r})=0.\nonumber
\end{eqnarray}
Using this identity, 
one can get that the energy $E_\ep(u_\ep)$ decouples as follows
\begin{eqnarray}
\label{splitener}
&&E_\ep(u_\ep)=E_\ep(\eta_\ep)+G_\ep(v_\ep)\nonumber\\
&&\;\;+2\int_{\cd} |\eta_\ep|^2 (\nab S_\ep - \bO\times {\bf r}) \cdot
(iv_\ep,\nabla v_\ep), 
\end{eqnarray}
where
$$G_\ep(v_\ep)=\int_{\cd} |\eta_\ep|^2|\nabla
v_\ep|^2+{{|\eta_\ep|^4}\over{2\ep^2}}(1-|v_\ep|^2)^2.$$ 
This decoupling was used in \cite{S} in the case of a disc
where $\nabla S_\ep=0$. 

\subsection{Estimate on the energy}
We now estimate the terms in (\ref{splitener}). The first
term $E_\ep(\eta_\ep)$ is a constant only depending on $\ep$, and not
on the solution type, that is with or without vortices. The
second term gives a contribution coming from the vortices and the third
term is due to the vortices and rotation.

We use the analysis of vortices developed in 
\cite{BBH} and later in \cite{BR,S1,S}. 
Let $\cd'_\ep=D \setminus \{x,\ \dist (x,\partial \cd)\leq
\ep^\beta\}$, with $\beta <1$. Then in $\cd'_\ep$ it is possible to
define vortices for $v_\ep$ in the following way: there exist balls
$B_i=B(p_i,\ep^{\beta'})$ where $p_i$ are points in $\cd'_\ep$ at mutual
distance bigger than $8\ep^{\beta'}$ and $\beta'>\beta$, such that
$|v_\ep|\geq 1/2$ in $\cd'_\ep\setminus\cup_i B_i$. Moreover,
the degree $d_i=deg(v_\ep/|v_\ep|,\partial B_i)$ is not zero and there is an
estimate of the energy of $v_\ep$ in each ball $B_i$. This analysis
means that vortices are in fact defined in the balls where $v_\ep$ is less
than $1/2$ and has a non zero degree. This allows us to compute
$G_\ep(v_\ep)$ for which only the gradient term in the vortex balls will
give a contribution: each vortex gives a contribution in the amount of 
 $2\pi \lep$ due to its degree and a contribution of lower order  
which comes from the interaction  with the other vortices. Moreover,
$|\eta_\ep|^2$ is almost $a$:  
\begin{eqnarray}
\label{G}
&&G_\ep(v)=2\pi {\lep}\sum_i |d_i| a(p_i) \nonumber\\
&&\;\;-2\pi \sum_{i\neq j} d_i d_j
a(p_i)a(p_j){\ln} |p_i-p_j|+O(1). 
\end{eqnarray}
In order to estimate the third term in (\ref{splitener}) we let $X_\ep$
be the solution of $a({\bf r})(\nab S_0-\bO\times {\bf r})=\O\np X_\ep$
which is zero on the boundary of $\cd_\ep'$. That is $X_\ep$ solves
(\ref{xilim}) but with zero boundary data on $\partial\cd_\ep'$ instead of
$\partial\cd$. Hence $X_\ep$ converges to $\xi$ the solution of (\ref{xilim}). 
 An integration by part on the last term of (\ref{splitener}) using the
definition of $X_\ep$ and the definition of the degree of $v_\ep$ on
vortex balls and the fact that the higher order term comes from an
integration on the vortex balls yields 
\begin{eqnarray}\nonumber
&&\int_{\cd} |\eta_\ep|^2 (\nabla S_\ep-\bO\times {\bf r}) \cdot
(iv,\nabla v)\\ &&
\quad
=\int_{\cd'_\ep\setminus \cup_i B_i} \bO  \cdot
(iv,dX_\ep\times \nabla v)(1+o(1))\nonumber\\
&& \quad
=\sum_i 2\pi\O d_i X_\ep(p_i)(1+o(1))\nonumber  \\&&
=
\sum_i
{-\pi\O d_i\over{1+\la^2}}  (\a-|x_i|^2-\la^2 |y_i|^2)^2
(1+o(1))\,. \label{xpi}  
\end{eqnarray}
Finally, one can derive from (\ref{splitener})-(\ref{G})-(\ref{xpi}) 
an asymptotic development of the energy for a solution with vortices. 
\begin{eqnarray}
&&E_\ep(u_\ep)-E_\ep(\eta_\ep)\simeq
2\pi \lep\sum_i |d_i|a(p_i)\nonumber\\
\nonumber
&&\quad - 2 \pi \sum_{i\neq j} d_i d_j
a(p_i)a(p_j)\ln |p_i-p_j|\\
&& \;
-{\pi\O\over{1+\la^2}} \sum_i d_i\Bigl (\a-|x_i|^2-\la^2 |y_i|^2
\Bigr)^2. \label{dec}
\end{eqnarray}
Note
that the minimal energy for solutions without vortices in $\cd'_\ep$ is
$E_\ep(\eta_\ep)+O(\ep\lep)$: it is not exactly $E_\ep(\eta_\ep)$ since
$\eta_\ep$ is a minimizer without the constraint $\|\eta_\ep\|=1$,
but it almost equals to $E_\ep(\eta_\ep)$ since $\int_{\cd} a=1$ and 
$|\eta_\ep|^2$ approaches $a$ asymptotically.

\section{Critical velocities}
\subsection{Critical velocity for the existence of one vortex}
Let $u_\ep$ be a minimizer of $(P)$ with one vortex at a point $p$ in
$\cd$ with coordinates $(x,y)$ and let  $\Delta E_\ep$ be the difference
between $E_\ep(u_\ep)$ and the energy of a solution without vortex
($E_\ep(\eta_\ep)+O(\ep\lep)$): 
\begin{eqnarray}\label{deltaEellipse}
&&
\Delta E_\ep =2\pi \Bigl (\a -x^2-\la^2y^2\Bigr )\cdot\nonumber\\
&&
\Bigl ( \lep -{{\O
}\over {1+\la^2}}  (\a -x^2-\la^2y^2  )\Bigr )(1+o(1)). 
\end{eqnarray}
This expression has been obtained by Svidzinsky and Fetter \cite{SF}
using a different method.  
The form of $\Delta E_\ep$ allows the computation of two critical velocities
$\O_s$ and $\O_1$ for the existence of vortices: $\O_s$ is the velocity
for which the solution with one vortex starts to be locally stable and
$\O_1$ for which it starts to be globally stable. For $\O < \O_s$,
$\Delta E_\ep$ is a decreasing function of $|p|$, the position of the
vortex; $|p|=0$ is a local maximum of $\Delta E_\ep$. For $\O_s < \O <
\O_1$, $|p|=0$ is a local minimum for $\Delta E_\ep$. Note that $\Delta
E_\ep(p\in\partial D)=0$ and $\Delta E_\ep(|p|=0)>0$. 

For $\O > \O_1$, $|p|=0$ is the global minimum for $\Delta E_\ep$. : 
\begin{eqnarray}
\nonumber
&&\O_s= {{1+\la^2}\over {2\a}}\lep ={{1+\la^2}\over{4\sqrt \la}}\sqrt{2\pi}
\lep  \; ,\\
\nonumber
&&\O_1= {{1+\la^2}\over \a}\lep ={{1+\la^2}\over{2\sqrt
\la}}\sqrt{2\pi}\lep
\end{eqnarray}
that is
$$\too_s={{1+\la^2}\over{4\sqrt
\la}}\omega\sqrt{{\pi\hb^2}\over{Ngm}}\ln \Bigl ( {{Ngm}\over{\hb^2}}
\Bigr ) ^{1/2} \; , $$
$$ \too_1= {{1+\la^2}\over{2\sqrt
\la}}\omega\sqrt{{\pi\hb^2}\over{Ngm}}\ln \Bigl ( {{Ngm}\over{\hb^2}}
\Bigr ) ^{1/2}.$$ 
Note that Castin-Dum \cite{CD} for the case
$\la=1$ find $\too_1=\omega\sqrt{{(\pi\hb^2)}/{(Ngm)}}\ln \Bigl ( (C/ \sqrt{\pi}) {{(Ngm)}/{\hb^2}}
\Bigr ) ^{1/2}$, with $C\simeq 1.8$, hence  $C/\sqrt{\pi} \simeq 1$ which gives a value of $\too_1$ very close to ours. They also have
 $\O_1=2\O_s$ for the case
$\la=1$.  Increases in anisotropy yield
higher $\too_1$ as already noticed  in \cite{FCS}, but as $\la$ tends to
infinity, $\O_1$ becomes bigger than $1/\ep$ so that vortices cannot be
stabilized. 

It can be proved that there exists $k_\ep$ which tends to zero with $\ep$
such that for $\O <\O_1-k_\ep$, the minimizer of $E_\ep$ has no vortex
and for $\O >\O_1+k_\ep$, there exists a minimizer with a
vortex. Moreover, for $\O_1+k_\ep < \O <\O_1+O(1)$, any minimizer has
one vortex of degree 1 tending to the origin. The proof consists in
constructing a test function with a vortex at the origin and computing
the energy of this test function. This yields an upper bound for the
energy. The lower bound relies on estimates for $G_\ep(v)$ from
\cite{BBH} and \cite{S1,S}.

\subsection{Critical velocity for $n$ vortices}
Similarly, one can compute $\O_n$, 
the critical velocity for the existence of  $n$ vortices. For this
purpose, one can prove that as $\ep$ goes to 0, the vortices tend to the
origin and they are of degree 1 that is $d_i=1$. This is similar to
\cite{S1}, \cite{S}. The test function consists in putting the $n$
vortices on a polygon  centered at the
origin of size $1/\sqrt \O$ in $x$ and $1/\la \sqrt \O$ in $y$. 
 More precisely, we let $\tilde p_i$ with coordinates $(\tilde x_i,\tilde
y_i)$ be such that $\tilde x_i=x_i\sqrt \O$ and $\tilde
y_i=\la y_i\sqrt \O$. This allows us to estimate the energy of a solution with $n$
vortices centered at $\tilde p_i$ from (\ref{splitener}),(\ref{G}),(\ref{xpi}): 
\begin{eqnarray}
&&E_\ep(u)=E_\ep(\eta_\ep)+2\pi n\a(\lep-{1\over {1+\la^2}}\O
\a)\nonumber \\
&&\label{se}
 +{\pi}(n^2-n) \a^2\ln \O
+w(\tilde p_1,...,\tilde p_n)+C_n+o(1)
\end{eqnarray}
where $C_n$ is a constant that depends on $n$ and $\la$ and
\begin{eqnarray}\nonumber
&&
w(\tilde p_1,..., \tilde p_n)=-\pi \a^2\sum_{i\neq j} \ln \Bigl (|\tilde x_i-\tilde
x_j|^2+{{|\tilde y_i-\tilde
y_j|^2}\over \la^2}\Bigr )\\ 
&&\quad
+2\pi \a\sum_{i} (\tilde x_i^2+ \tilde y_i^2)\Bigl ({2\over
{1+\la^2}}-{{\lep}\over \O \a}\Bigr ).\label{w} 
\end{eqnarray}
Recall that $\a^2=2\la/\pi$. For fixed $\la$, $w$ is of order 1, hence is of lower order than the previous terms.
 Then the critical velocity for the existence of  $n$ vortices can be computed
from (\ref{se})
\begin{eqnarray}
\label{On}
&&\O_n=(1+\la^2)\sqrt{\pi\over {2\la}}\lep\nonumber\\
&&\;\;+{{(1+\la^2)}\over 2}(n-1)\ln
\Bigl ({{(1+\la^2)}\sqrt{\pi\over {2\la}}}\lep\Bigr ) ,
\end{eqnarray}
and the critical velocity in the original parameters is
$\too_n$ 
\begin{eqnarray}\nonumber
&&
\too_n={{(1+\la^2)}\over{2
}}\omega\Bigl (\sqrt{{\pi\hb^2}\over{Ngm\la}}\ln \Bigl ( {{Ngm}\over{\hb^2}}
\Bigr ) ^{1/2}  
\\\nonumber
&&\;\;+ (n-1)
\sqrt{{ \hb^2}\over{2Ngm}}\ln\Bigl(
{{1+\la^2}\over {2\sqrt \la}} \sqrt{2 \pi}\ln(
{{Ngm}\over{\hb^2}}  ) ^{1/2}\Bigr)\Bigr).
\end{eqnarray}

\subsection{Location of vortices}
Once $\O$ is close to $\O_n$, the location of the vortices is
characterized by the configuration of points $\{\tilde 
p_i\}$ which minimizes  the function $w$ given by (\ref{w}).
In non-dimensionalized variables, the points are given by $R p_i/
\sqrt{2\pi\lep}$. For convenience, we define
$$\rho= \sqrt{{2\pi}\over \la} \Bigl ({2 \over
{(1+\la^2)}}-{\lep\over{\O\a}}\Bigr ).$$ 
Note that given the value of $\O_n$ in (\ref{On}), to leading order,
$\rho$ is equal to $\sqrt{{2\pi} / \la}/(1+\la^2)$. We use the value of
$\a$ and $\rho$ to get a simplified expression for $w$: 

\begin{eqnarray}\nonumber
&&
w(\tilde p_1,..., \tilde p_n)=-2\la \Bigl ( \sum_{i\neq j} \ln \Bigl (|\tilde x_i-\tilde
x_j|^2+{{|\tilde y_i-\tilde
y_j|^2}\over \la^2}\Bigr )\\ 
&&+\rho\sum_{i} (\tilde x_i^2+ \tilde y_i^2) \Bigr).
\label{w2} 
\end{eqnarray}
The critical points of $w$, and thus the vortex positions,
 satisfy:
\begin{eqnarray}
\label{nece}
{\rho} \tilde{x}_i = 
 \sum_{j\neq i} {\lambda^2 (\tilde{x}_i-\tilde{x}_j) \over \lambda^2
|\tilde{x}_i-\tilde{x}_j|^2+|\tilde{y}_i-\tilde{y}_j|^2 }\;,\\[-0.2cm]
{\rho} \tilde{y}_i = 
 \sum_{j\neq i} {\tilde{y}_i-\tilde{y}_j \over \lambda^2
|\tilde{x}_i-\tilde{x}_j|^2+|\tilde{y}_i-\tilde{y}_j|^2 }\;.
\end{eqnarray}
An immediate observation is that 
\begin{equation}
\label{hyperplane}
\sum_i \tilde{x}_i = \sum_i \tilde{y}_i = 0\;. 
\end{equation}
By multiplying the equations with $\tilde{x}_i$ and $\tilde{y}_i$
respectively and adding the results together, one can obtain
\begin{equation}
\label{hypersphere}
\sum_i (\tilde{x}^2_i+ \tilde{y}^2_i) = n(n-1)
 /(2 \rho) 
\; .
\end{equation}
Similarly, multiplying the equations with $\tilde{y}_i$ and
$-\lambda^2\tilde{x}_i$ respectively and adding the results together, one gets
\begin{equation}
\label{orthog}
\rho(1-\la^2)\sum_i \tilde{x}_i \tilde{y}_i = 0
\; .
\end{equation}
Unlike equations (\ref{hyperplane},\ref{hypersphere}) where the
dependence on $\la$ is implicit, the equation (\ref{orthog}) leads
to a property
\begin{equation}
\label{orthog2}
\sum_i \tilde{x}_i \tilde{y}_i = 0 \; , \quad \mbox{ for }\; \la\neq 1\; .
\end{equation}
The above observations lead to more precise predictions
on the location of vortices.

For instance, in the case $n=2$ we get $\tilde{x}_1=-\tilde{x}_2$
and $\tilde{y}_1=-\tilde{y}_2$. For $\lambda=1$, we have an
infinite set of solutions consisting in two points 
on the circle $\tilde{x}_i^2+\tilde{y}_i^2=1/\rho$, 
symmetric with respect to
the origin. For $\lambda \neq 1$, (\ref{orthog2}) leads to
$\tilde{x}_i\tilde{y}_i=0$ for $i=1,2$, and we have a pair of solutions with
either $\tilde{x}_i=0$, $\tilde{y}_i=\pm \sqrt{1/2\rho}$, or
$\tilde{y}_i=0$, $\tilde{x}_i=\pm \sqrt{1/2\rho}$. 
Checking the corresponding values of $w$, 
we get that for $\lambda \neq 1$, the minimizer of $w$ 
corresponds to having both
vortices staying on the long-axis of the 
ellipse in the original scaling (that is on the $x$ axis if $\la>1$ and on the $y$ axis if $\la<1$). The estimate of the location
is in agreement with the numerical solutions given later.

For the case $n=3$, we also get that the 3 vortices are on the long-axis of the ellipse: one centered at the origin whereas the other pair   stays symmetrically on the long-axis with $\tilde{x}_i=\pm \sqrt{3/2\rho}$ if $\la>1$.
 Similar discussions can be carried out for other values of $n$.
In general, for the case of $n$ vortices, we let $R_1^2=n(n-1)/2\rho$ and
$\hat x_i= R_1\tilde x_i$, $\hat y_i= R_1\tilde y_i$. It follows from
(\ref{hypersphere}) that the points $\{\tilde p_i\}$ are localized by the
minimum  of
$$-\sum_{i\neq j}\log\Bigr( |\hat x_i-\hat x_j|^2+{{|\hat y_i-\hat y_j|^2}
\over \la^2} \Bigl ).$$ 
under the constraint $\sum_i \hat{x}_i ^2+\hat{y}_i^2 =1$.
It is reasonable to expect that this formula will
lead to a vortex array as observed in \cite{FCS2}. Note that
the minimizer $\{\hat p_i= R_1 \tilde p_i\}$ has no
explicit dependence on $\O$, thus, we expect for a given
$\la$, the vortex configuration to be of similar structure for values of
$\O$ close to $\O_n$. 
On the other hand, for a given $n$ and $\O$, we expect, however, 
that for sufficiently large $\la$, that 
is for highly anisotropic traps, the minimizer of $w$ is
given by the collinear solutions with vortices all located on the
long axis of the elliptical trap.

To our knowledge, theoretical investigation had been restricted to the
critical velocity for nucleation of one vortex. Here, we are able to
deal with the case of multiple vortices and to
precisely characterize the location of the vortices.
Our estimates are also consistent with the numerical results given later.

\section{Evolution equation and numerical schemes}
To numerically compute the energy minimizers of (\ref{be}),
we notice that the energy in (\ref{be}) can be rewritten as
\beq\label{bhk}
\int_{\cd} \left\{|(\nabla  - i \ba)u |^2
+{1 \over  2\ep^2}(a_\ep({\bf r})-|u|^2)^2 \right\} +c_\ep
\eeq 
where $a_\ep({\bf r})=a({\bf r}) -\ep^2\O^2r^2$,
$$
\ba=\left(
\begin{array}{c}
y\\
-x\end{array}\right)\bO\;, \;\;\mbox{ and }\;\;
c_\ep=\int_{\cd} \{{1\over  2\ep^2} 
(a^2({\bf r})-a^2_\ep({\bf r}) )
\}.$$
The above formulation of the energy has a striking similarity with the
high-kappa high-field Ginzburg-Landau energy  \cite{CDGP} with a
variable coefficient \cite{DGP2}.

\subsection{Evolution in the imaginary time}
To numerically compute the minimizers of (\ref{be}), we 
consider the time-dependent equation  in the imaginary time:
\begin{equation}\label{tdgp}
\frac{\partial u}{\partial t} -  \left(\nabla - i \ba\right)^2u
   + {1\over \ep^2}|u|^2u - \frac{a_\ep(\br) }{\ep^2} 
u  = \mu_\ep(u) u 
 \; 
\end{equation}
in $\cd$ with initial condition
$u(\br,0)=u_0(\br)$ in $\cd$ and
boundary condition $u=0$ on $\partial \cd$. 
 Here, $\mu_\ep (u)$ denotes
the Lagrange multiplier.
Assume that $u_0$ satisfies the constraint $\|u_0\| = 1$.
Then, by taking
$$\mu_\ep(u) = 
\int_{\cd} \left\{  \left|\left(\nabla - i \ba\right)u\right|^2
+ {1\over \ep^2} |u|^4-
 {a_\ep(\br)\over \ep^2} |u|^2 \right\} d\,\cd\;,
$$
we get
$$
\frac{d}{d t}
\left(\int_{\cd} |u|^2 d\cd -1\right)
-  \mu_\ep(u) \left( \int_{\cd} |u|^2 d\cd -1\right) = 0 \; .
$$
Thus, the constraint
$\int_{\cd} |u|^2 = 1$ is ensured at all time.
Moreover,
using $(u, u_t)=0$, we get the energy estimate:
$$
\frac{1}{2}\frac{d}{d t} \ce_\ep(u) + \|\frac{\partial u}{\partial t}\|^2
= 0\;.
$$

Thus, we easily get that for any $(0,T)$, if $u_0\in 
 H^1_0(\cd)$ and $\|u_0\| = 1$,  
there exists a unique strong solution
$u$ of (\ref{tdgp}) satisfying the constraint $\|u\| = 1$.
Using argument similar to that in \cite{LS,LD}, we may also get that
as $t\to \infty$, $u$ approaches to a steady state solution
which is a critical point of the energy. 
Well-posedness for $L^2$ initial data may also be be obtained.

\subsection{Numerical schemes}
There are various ways to solve the time-dependent Gross-Pitaevskii
equations, see for example \cite{CD} or \cite{FCS}. We  take the
advantage of the striking similarity with the high-kappa high-field
time-dependent Ginzburg-Landau equations \cite{DG}, and adapt a code
developed in \cite{DG,DGP,DGP2}. Spatially, we use a standard
finite element approximation, see \cite{DGP,DGP2} for details. Here, we
focus on the time-discretization and the treatment of constraint.
It has been observed that there are some steady states exhibiting
meta-stability, thus it is important to 
get asymptotically stable schemes for large time which 
in general requiring the use of implicit schemes with no 
limitations on the time step size.

Let $\{u_n\}$ be approximate solutions of $\{u(t_n)\}$
at discrete time $\{t_n\}$ with time-step
$\Delta t_n=t_n-t_{n-1}$. 
We discuss two time-discretization schemes and also some 
results of numerical experiments. 
\paragraph{A first order backward-Euler 
in time discretization}
Given $u_{n-1}$, we first solve for $u^*$:
\begin{eqnarray}
\label{eq:dis1}
&&\frac{u^*-u_{n-1}}{\Delta t_n}
-\left(\nabla - i \ba\right)^2u^*
   -\mu(u_{n-1}) u^* \nonumber\\
&&\quad\quad + {1\over \ep^2}|u^*|^2u^*- 
\frac{1}{\ep^2} a_\ep  u^*= 0
\end{eqnarray}
Then, we apply the projection $u_{n}=u^*/\|u^*\|$. 
Both the backward Euler step and the projection step gives only
first order in time accuracy.
\paragraph{A norm-preserving, energy-decreasing second order
scheme}
For any $u$, $v$ and their complex conjugate
$\bar{u}$, $\bar{v}$, we let  $f(u,v)=(|u|^2+|v|^2)(u+v)/2$ which
satisfies
$f(u,v){(\bar{u}-\bar{v})}=(|u|^4-|v|^4)/2$. 
 Given $u_{n-1}$, we first solve for $u^*$:
\begin{eqnarray}
&&\frac{2(u^*-u_{n-1})}{\Delta t_n}
-\left(\nabla - i \ba\right)^2u^*    -\nu(u^*) u^* \nonumber \\
&& \quad
+ {1\over \ep^2}f(2u^*-u_{n-1},u_{n-1}) - 
\frac{1}{\ep^2} a_\ep  u^*= 0
\label{eq:dis2}
\end{eqnarray}
where $\nu(u^*)$ is given by
\begin{eqnarray}
&&\nu(u^*)\int_{\cd} |u^*|^2  =
\int_{\cd} \left\{  \left|\left(\nabla - i \ba\right)u^*\right|^2\right\}
\nonumber\\
&&\quad\quad + \int\left\{ {1\over \ep^2} f(2u^*-u_{n-1},u_{n-1})\bar{u}^*-
 {a_\ep\over \ep^2} |u^*|^2 \right\} \; . \nonumber
\end{eqnarray}
Then we let  $u_n=2u^*-u_{n-1}$. 
 Taking the inner product of the equation (\ref{eq:dis2}) with 
$u^*$, we get $(u^*-u_{n-1}, u^*)=0$, which leads to 
$\|u_n\|^2=\|u_{n-1}\|^2 $. That is, the norm is preserved 
at each time step. 
Taking the inner product of the equation (\ref{eq:dis2}) with 
$u^*-u_{n-1}$, it is easy to get
$$ 2\frac{\|u_n-u_{n-1}\|^2}{\Delta t_n} + \ce_\ep(u_n) -
\ce_\ep(u_{n-1}) = 0 .$$
Thus, during the discrete time evolution, 
the energy decreases. This discrete scheme is second order in time
and unconditionally stable. It also preserves some essential features
of the continuous dynamic system, making it suitable for long time
integration and for studies of meta-stabilities of the solutions.

\subsection{Description of the numerical experiments}
We have used the above schemes to calculate various numerical solutions
for the parameter values $\ep=0.02$, $\lambda=1$ and $\lambda=1.5$.
The spatial finite element space is taken to be $C^0$ 
piecewise quadratic elements on triangular meshes. 
As we are mostly interested in the minimizers of the Gross-Pitaevskii
energy, the time-evolution is employed as a means of marching to
the steady state solutions. For this reason, we have used variable
time-steps in order to accelerate the convergence in time.
The nonlinear systems are solved by a Newton like methods at
each step. Though such a method is computational costly per step,
this drawback is offset by its unconditional asymptotic stability
for marching to the steady state.
We have also computed the solutions using refined meshes to ascertain the
numerical convergence. 

To obtain solutions for various velocities, we have used a number of
differential initial conditions. For example, we have used
$|u_0(\br)|^2=a({\bf r})$ for $\br \in \cd$ which serves as a good
approximation to the steady state solution, especially in the case
of vortex-free solutions. We note that for large values of $\O$, this
choice of initial condition can also lead to steady state solutions with
multiple vortices. Detailed solution branches are described in the next
section.  In addition, we have also used other initial conditions that 
manually {\em plant} vortices in the domain in order to find different 
solution branches. Finally, a continuation in the parameter $\O$ has
often been employed to follow a particular solution branch and to
compute the bifurcation diagrams. The continuation procedure also
provides a test for the local stability of the numerical solutions:
when one branch becomes unstable, the solution jumps onto a different
branch.

\section{Numerical results and bifurcation diagrams}
We now present some pictures of numerical solutions
and discuss the various solution branches.

\subsection{Description of solutions}
\begin{figure}[htb]
\vspace{-0.2in}
\centerline{\epsfxsize=0.6in\epsfbox{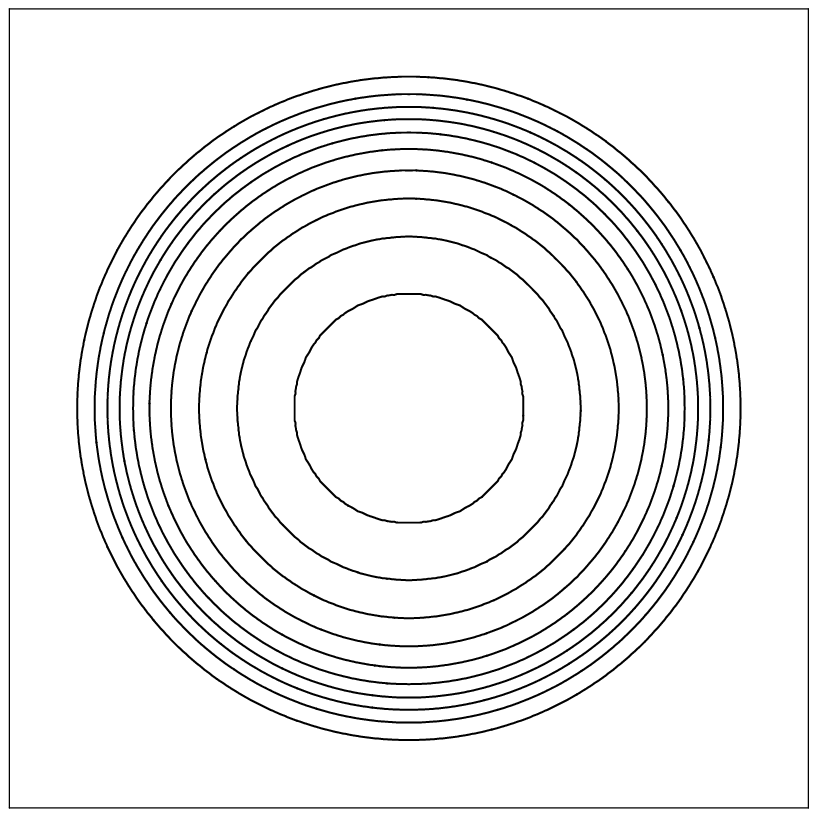}$\;$$\;$$\;$$\;$$\;$
            \epsfxsize=0.6in\epsfbox{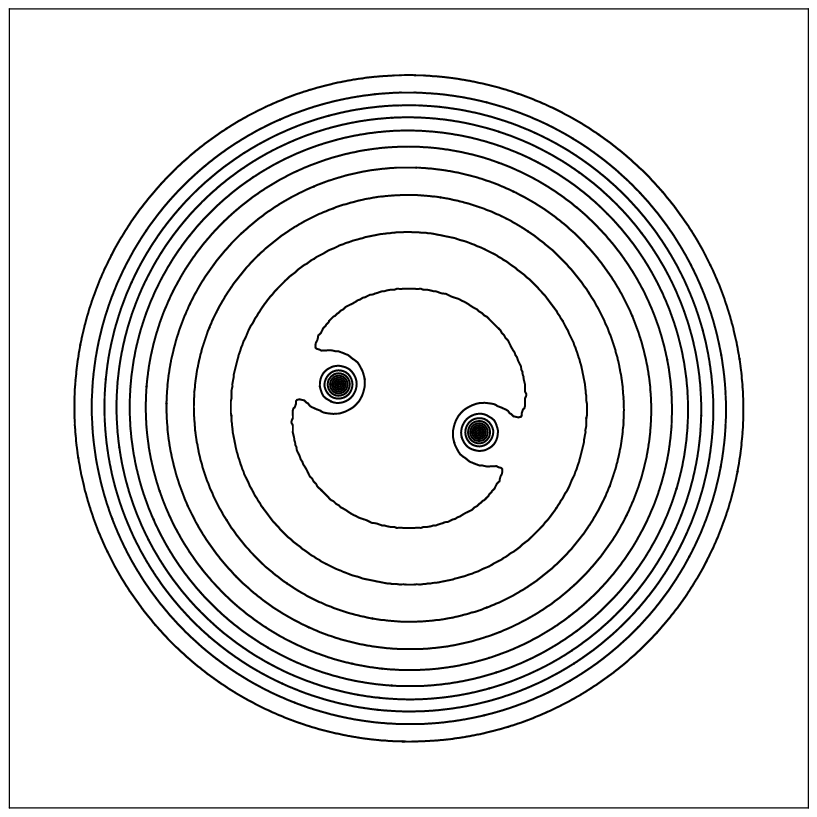}$\;$$\;$$\;$$\;$$\;$
	    \epsfxsize=0.6in\epsfbox{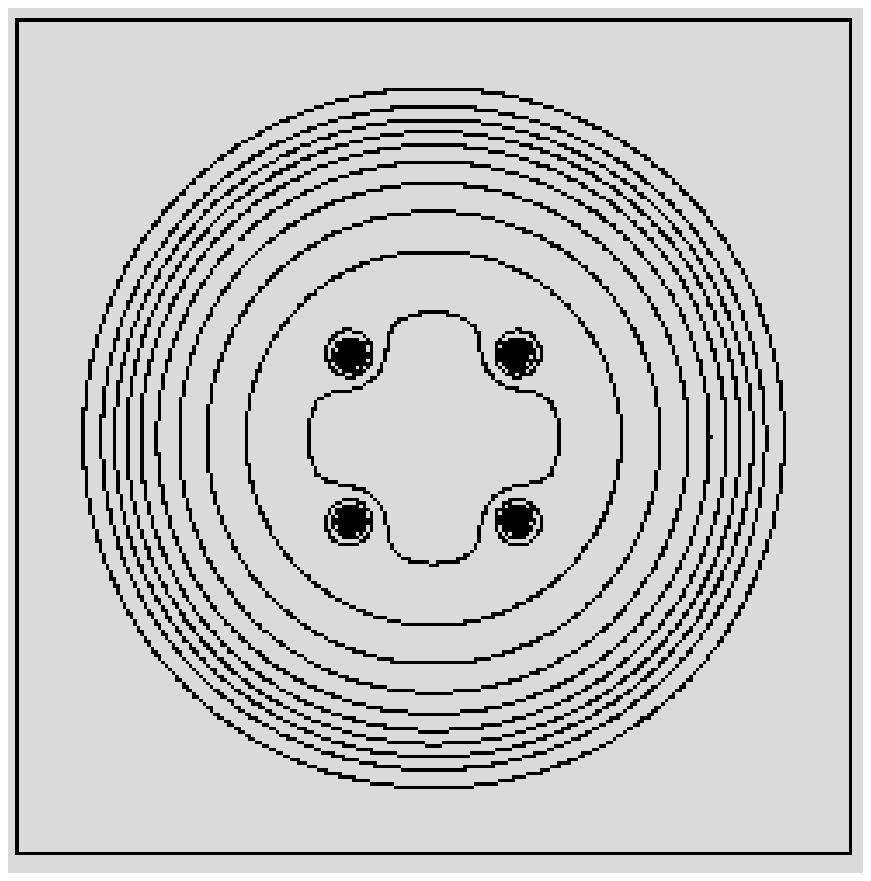}$\;$$\;$$\;$$\;$$\;$
	    \epsfxsize=0.6in\epsfbox{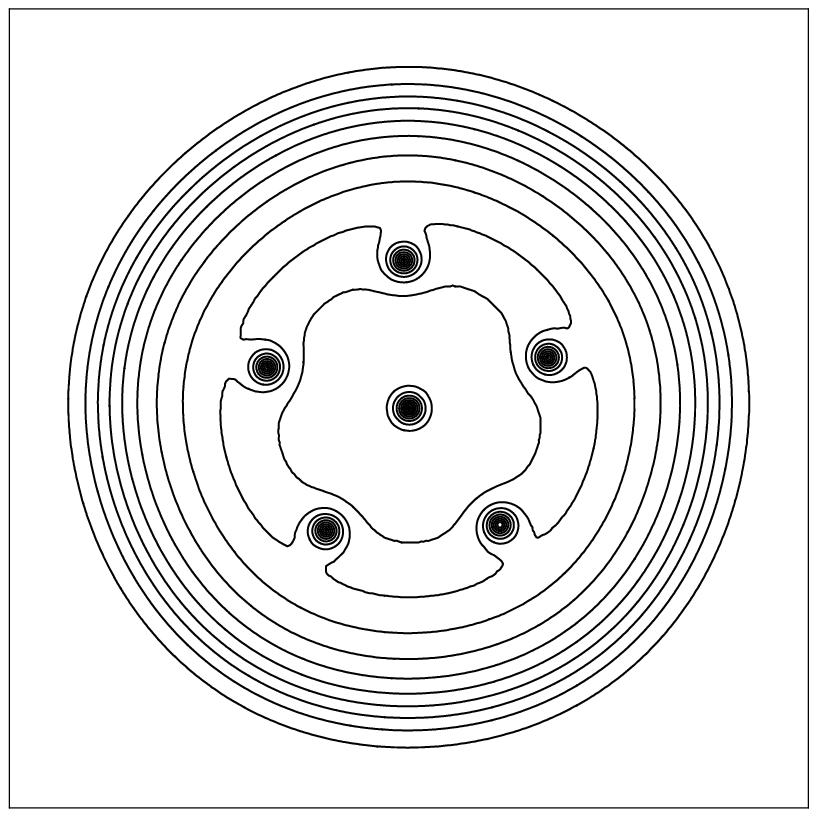}}
\vspace{-0.1in}
\centerline{\epsfxsize=0.6in\epsfbox{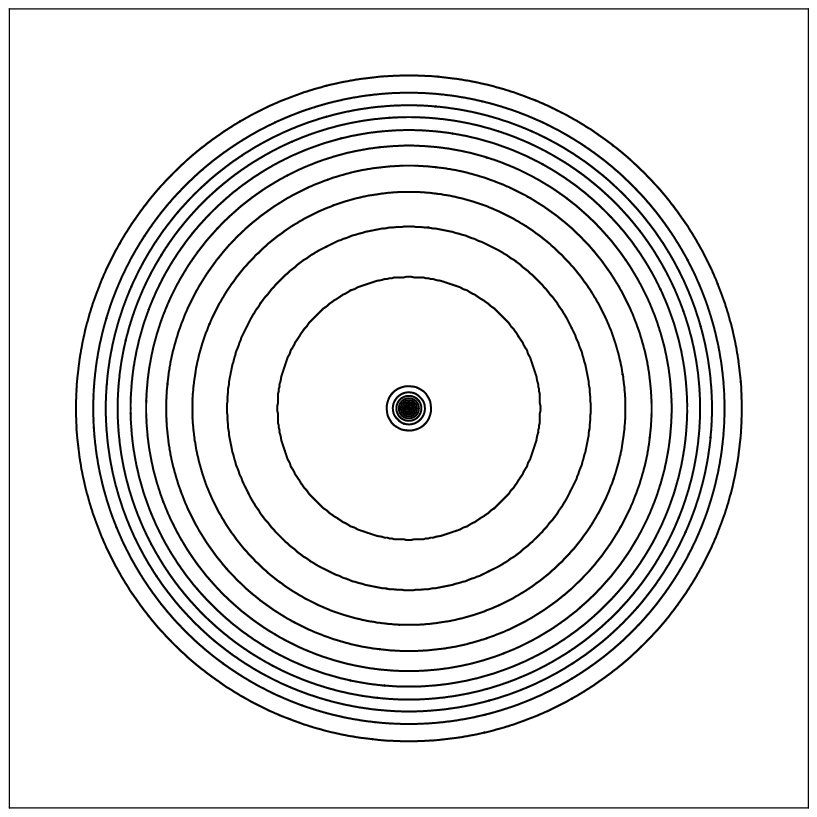}$\;$$\;$$\;$$\;$$\;$
	    \epsfxsize=0.6in\epsfbox{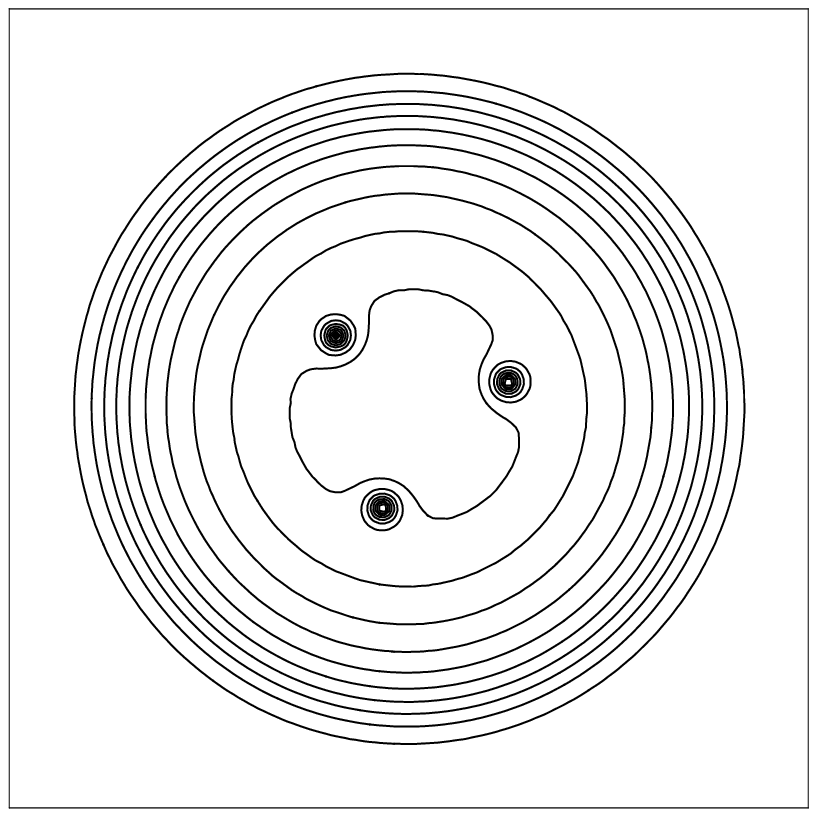}$\;$$\;$$\;$$\;$$\;$
	    \epsfxsize=0.6in\epsfbox{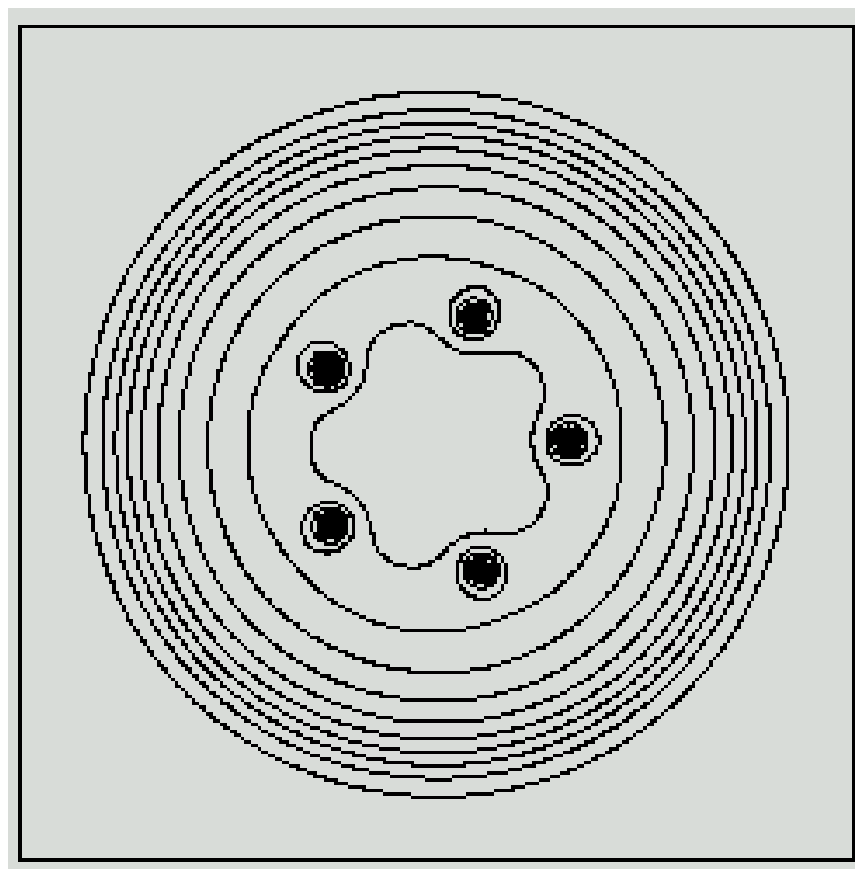}$\;$$\;$$\;$$\;$$\;$
	    \epsfxsize=0.6in\epsfbox{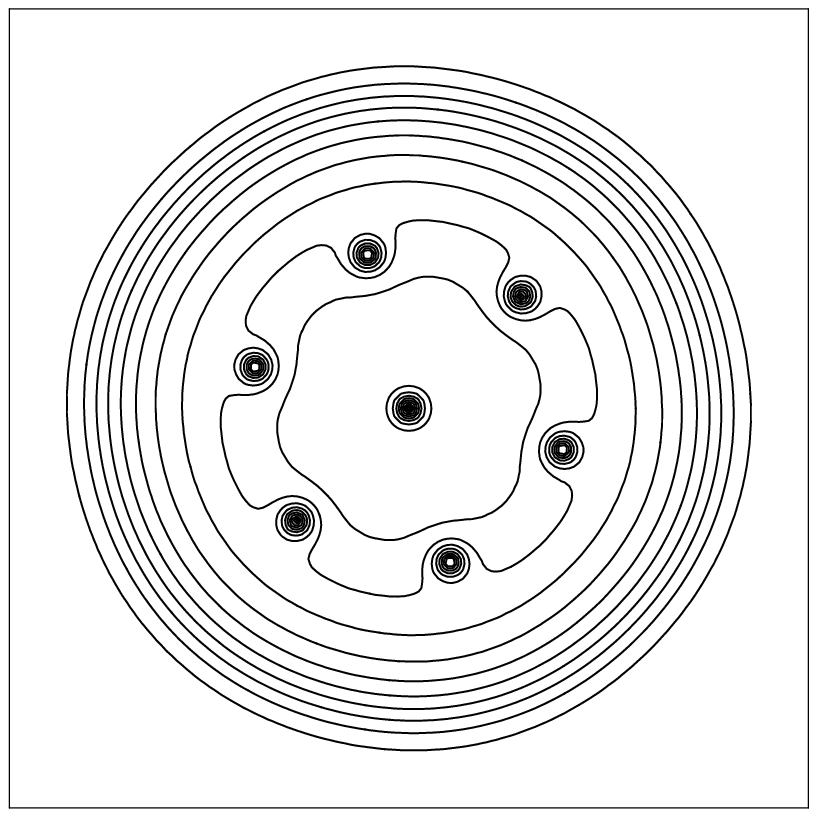}}
\caption{Contour plots of $|u|$ for $\Omega=0$ (1st column),
$15$ (2nd column) $17.5$ (3rd column) and $20$ (4th column).}\label{fig1}
\end{figure}

\paragraph{The case of a disc}
For any $\O$, there is a vortex free solution, which is close to
$a(x,y)$ except near the boundary. For $\O=0$, in addition to this
vortex free solution, there is also a one vortex solution, as
illustrated in  the first column of Figure \ref{fig1}.

\begin{figure}[htb]
\epsfxsize=1.5in\epsfbox{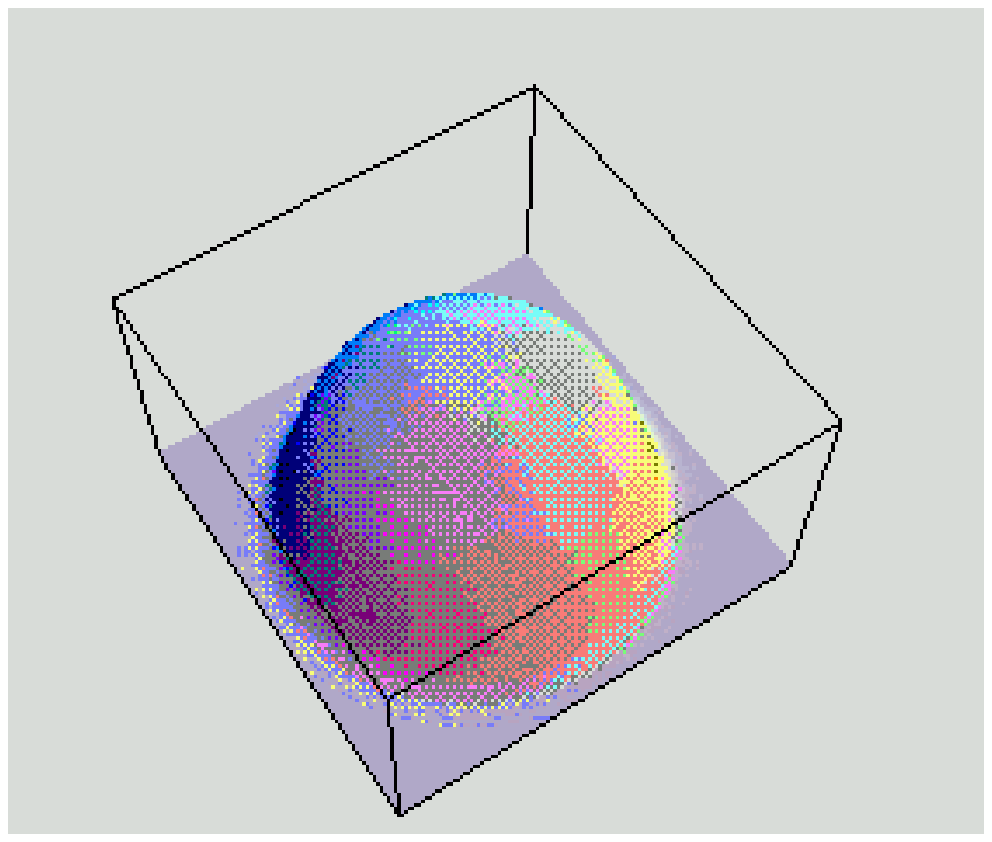}\hspace{-0.3in}
\epsfxsize=1.5in\epsfbox{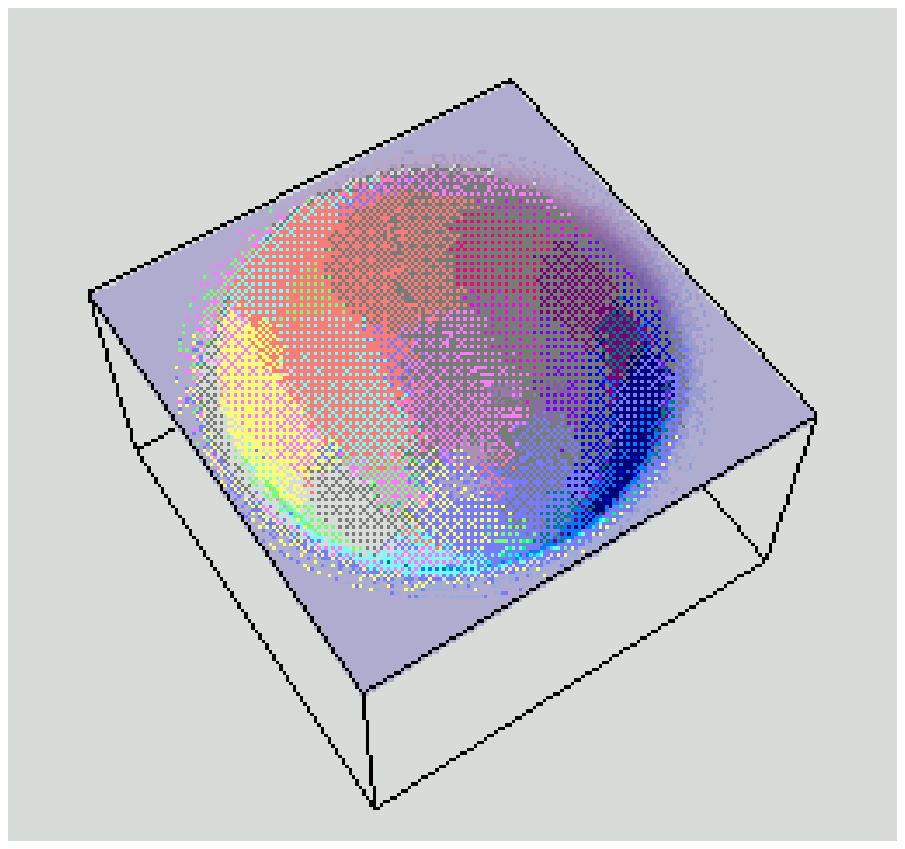}
\caption{Surface plots of $|u|$ for solutions at $\Omega=0$.}
\label{fig3D}
\end{figure}

\begin{figure}[htb]
\epsfxsize=1.5in\epsfbox{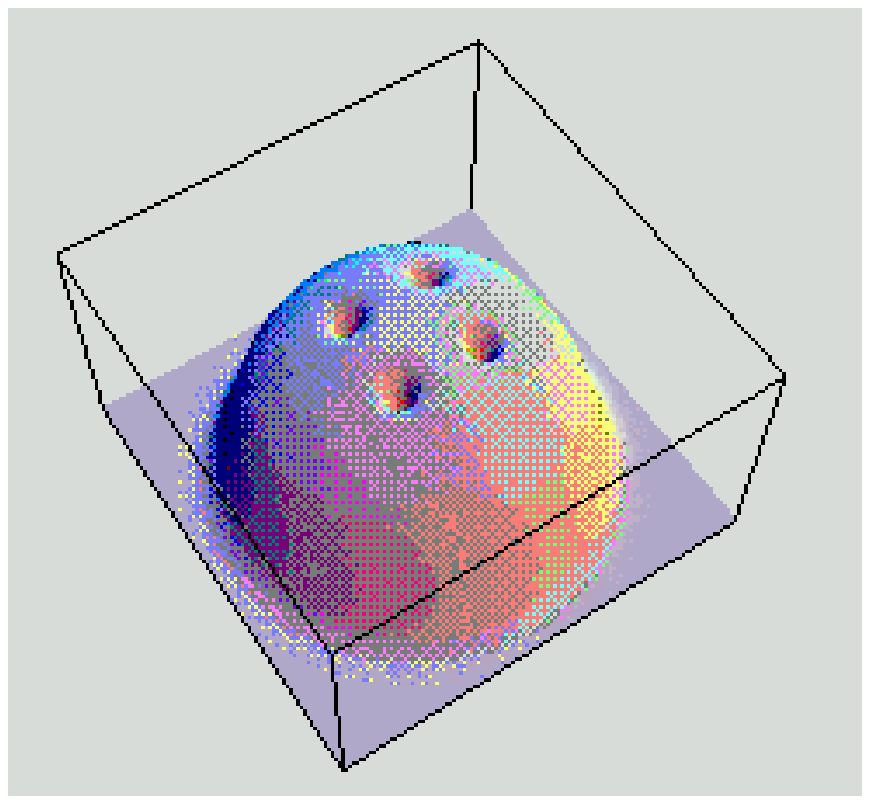}\hspace{-0.3in}
\epsfxsize=1.5in\epsfbox{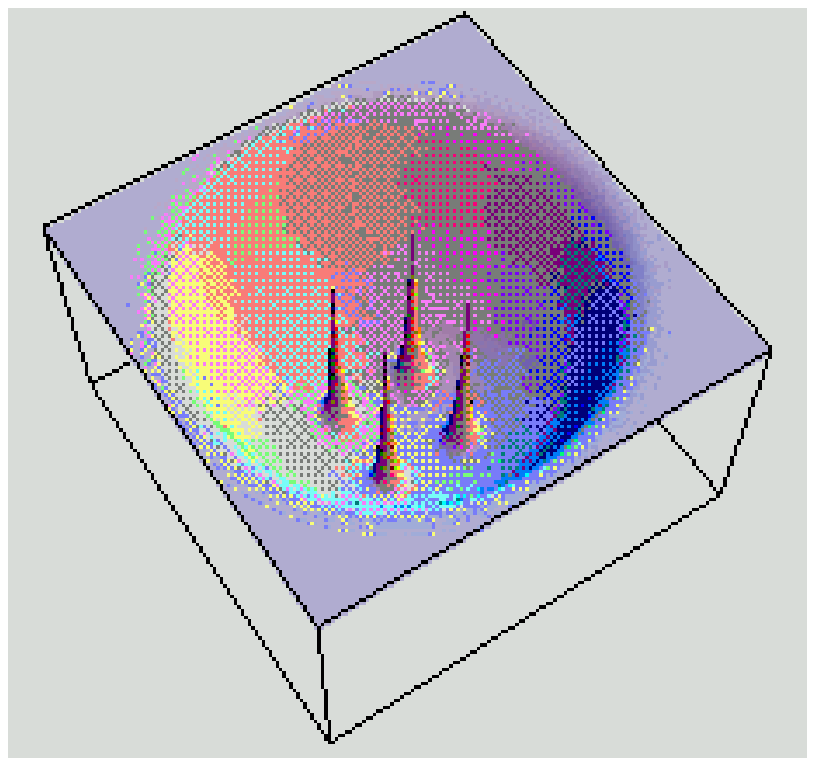}
\caption{Surface plots of $|u|$ for solutions at $\O=20$.}\label{fig3D2}
\end{figure}

For larger $\O$, solutions with multiple vortices 
are shown in the other columns of Figure \ref{fig1}.
For instance, solutions with 2 and 3 vortices for $\O=15$ are shown
in the second column,  solutions with 4 and 5
vortices for $\O=17.5$ are shown in the third column, 
 and solutions with 6 and 7 vortices for $\O=20$ 
are shown in the last column. 
 The parameter values for which the
single-vortex and multi-vortices solutions exist are to be presented in
the next section. Figures \ref{fig3D} and \ref{fig3D2}
provide
the surface plots (and better view) of $|u|$ for a vortex free solution 
at $\Omega=0$ and a solution at $\Omega=20$ with four vortices respectively.
Each solution has a top and bottom view, the paraboloid shapes 
are easy to visualize from the picture.

\begin{figure}[htb]
\vspace{-0.4in}
\centerline{\epsfxsize=1.in\epsfbox{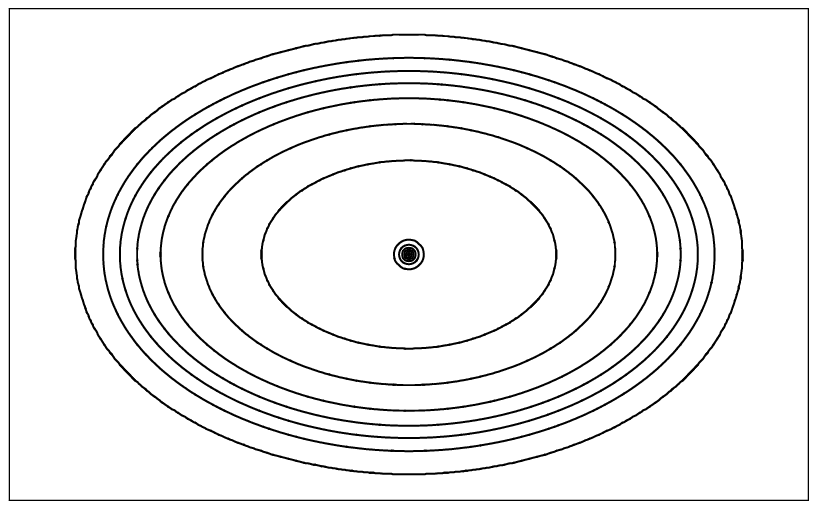}$\;$$\;$$\;$$\;$$\;$
\epsfxsize=1.in\epsfbox{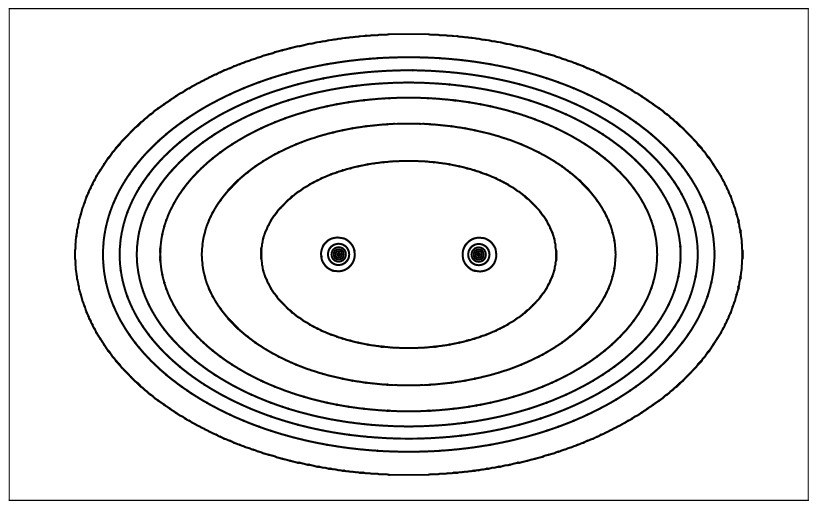}}
\vspace{-0.1in}
\centerline{\epsfxsize=1.in\epsfbox{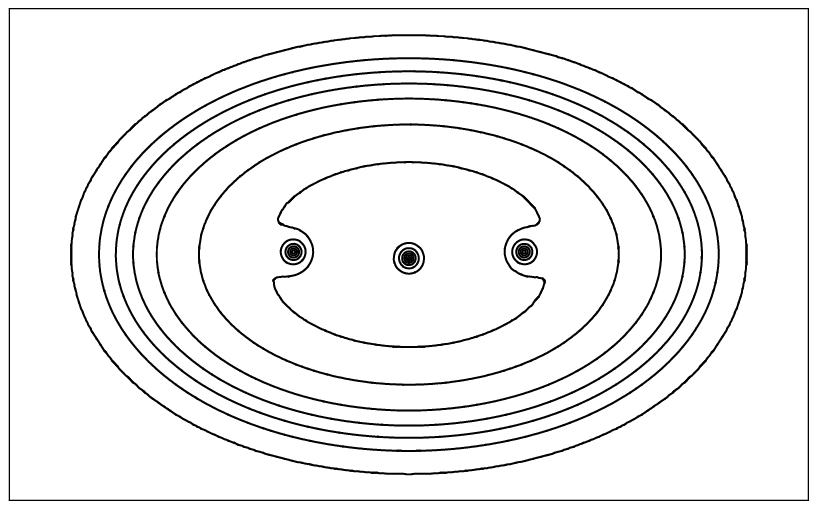}$\;$$\;$$\;$$\;$$\;$
\epsfxsize=1.in\epsfbox{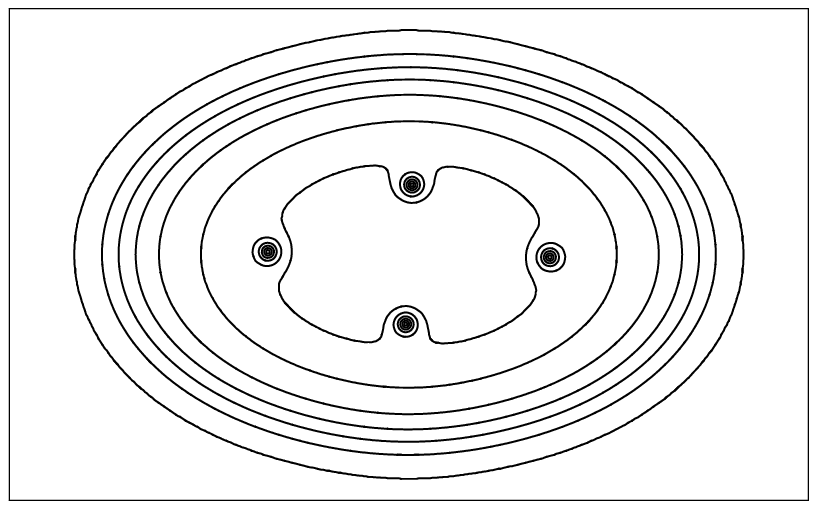}}
\vspace{0.2in}
\caption{Contour plots of $|u|$ for $\Omega=17.5, \lambda=1.5$}\label{fig5}
\end{figure}

\begin{figure}[htb]
\vspace{-0.3in}
\centerline{\epsfxsize=1.in\epsfbox{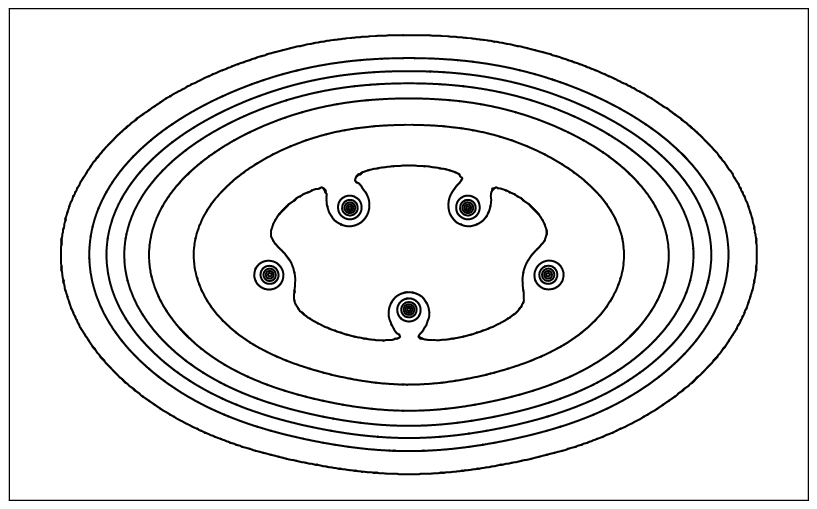}$\;$$\;$$\;$$\;$$\;$
\epsfxsize=1.in\epsfbox{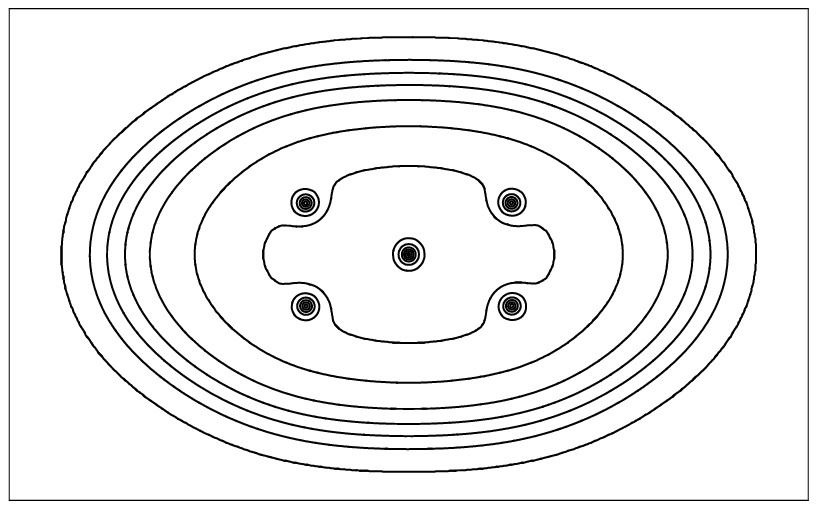}}
\vspace{-0.1in}
\centerline{\epsfxsize=1.in\epsfbox{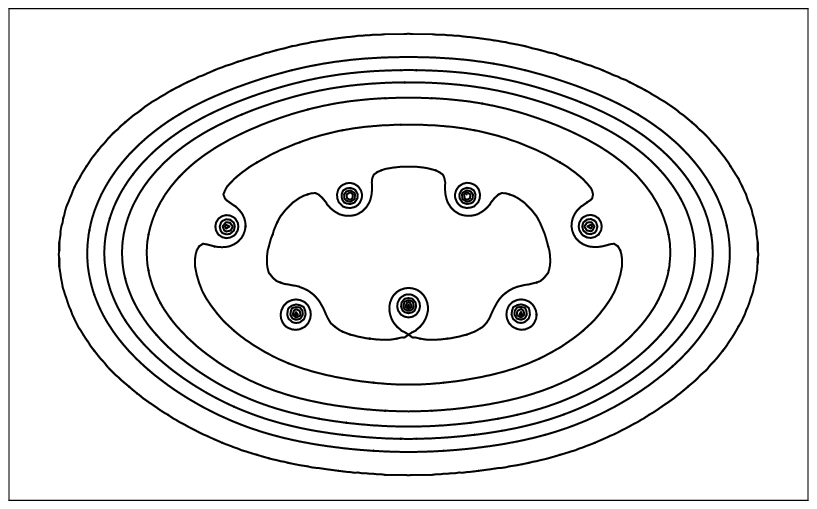}$\;$$\;$$\;$$\;$$\;$
\epsfxsize=1.in\epsfbox{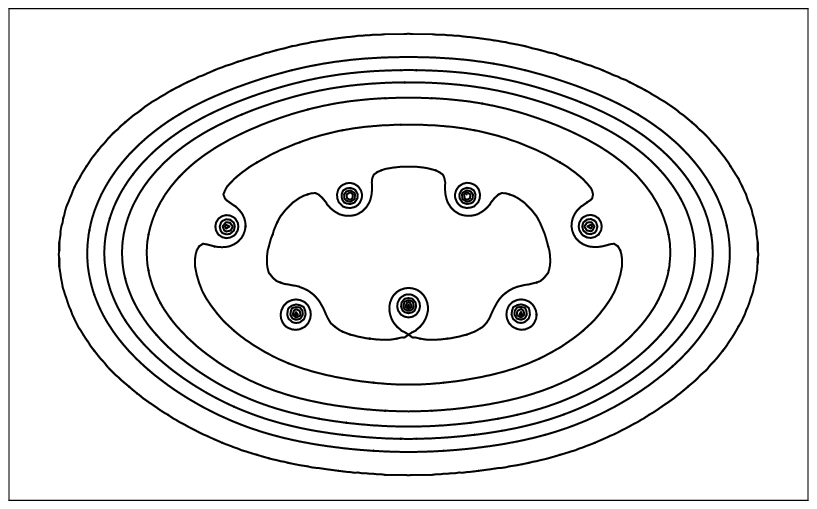}}
\vspace{0.25in}
\caption{Contour plots of $|u|$ for $\Omega=25, \lambda=1.5$}\label{fig6}
\end{figure}

\paragraph{The case of an ellipse}
We now present some solutions for an ellipse corresponding
to $\la=1.5$. In Figure \ref{fig5}, the contour plots of the magnitude of the
solution $|u|$ with $\O=17.5$ are drawn while in Figure 
\ref{fig6},  $\O=25$. For $\O=17.5$, it is interesting to compare the location of vortices with the analysis of section III.c: in this case $\rho=0.79$, the long axis is 0.9887 and we find for the location of the vortex $x=0.19$ for $n=2$ and $x=0.33$ for $n=3$. The picture shows that the vortex is about 1/5 of the long axis for 2 vortices and 1/3 for 3 vortices, which is consistent with the analysis. We believe that for a bigger number of vortices $n$, their location corresponds to the positions minimizing (\ref{w2}).


Note that there were two vortex configurations with five vortices for
$\O=25$,  among which, the non-symmetric configuration corresponds
to a solution with a lower free energy, though the difference in their
energy values is very small. 

\subsection{Branches of solutions}
An issue that we have studied is the existence of branches of $n$ vortex
solutions as $\O$ is varied and especially which one is the minimizer. 
\paragraph{The case of the disc}
For $\O=0$, the solution with lowest energy  is the vortex free
solution. We start with this solution as initial value for the time
dependent problem for a slightly bigger $\O$. This device allows us to
continue the branch of the vortex free solution as $\O$ is increased. We
find that the vortex free solution is obtained as the limit when $t$ is
large of the evolution equation up to $\O=19$. For $\O=19$, six vortices
are nucleated  from the boundary and eventually one vortex moves to the 
center and final configuration is similar to that in  Figure \ref{fig1}
(top right).
 Now if $\O$ is decreased from the value 19 using the 6 vortices
solutions as initial value, we see that this 6 vortices solution branch
exists down to $\O=16$ when it jumps to 4 vortices. If we decrease $\O$
further, then we stay on the 4 vortex branch down to $\O=13$ when it
drops to a 2 vortex branch. Similarly, if we increase $\O$ from 13, the
2 vortices solution will persist up to $\O=21$. 

As for the                                                            
one-vortex solution branch, it is computed by
planting a vortex-like function at the center in the initial
condition for $\O=10$, then the branch is computed by continuation
in $\O$ (both increasing and decreasing). It is interesting to observe
the fact that this branch extends all the way to $\O=0$, such 
 persistence of the
one-vortex solution even for the zero velocity 
has been elaborated by various authors, see  \cite{FCS} for instance.
Since implicit integration is used, we are able to indeed march to
to the steady state and to ascertain that this persistence is not due
to the metastability.
On the other hand if the vortex is planted away from the origin, it
disappears for small $\O$ as we will see later. 

\paragraph{The case of an ellipse}
The same kind of behaviour occurs for the vortex free solution branch
with $\la=1.5$: when $\O$ is increased from 0, we stay on the vortex
free branch up to $\O=22.5$ when the solution jumps to the 4 vortex
branch of Figure \ref{fig6}. If we decrease $\O$ starting from this 4
vortex branch, the solution will stay on it down to $\O=15$ when it
jumps to 2 vortices and at $\O=10$ it jumps to the vortex free
solution. Similarly to the case of the disc, if one vortex is planted at
the center at time 0, it will persist in time even down to $\O=0$. 

For $\O$ large enough, several vortices are nucleated at the same
time. Basically, we are counting on initial conditions and the round-off
errors to break the symmetry of the region and 
strong symmetry presence often makes symmetry breaking much harder
to achieve. For the disc case, we expect equal chance of vortex
nucleation from any point of the boundary. It turns out that
for $\O=19$, an unstable front produces oscillations of almost equal
magnitude, and spins off 6 vortices at the same time. 
Had the disturbance being unevenly distributed, then it would be
possible for some vortices to get spin off ahead of others, and 
due to the repulsion, the others may never have a chance to appear, 
thus we may see solutions with a smaller number of vortices.

\subsection{Energy diagrams}
\begin{figure}
\vspace{0.1in}
\centerline{\epsfxsize=2.8in\epsfbox{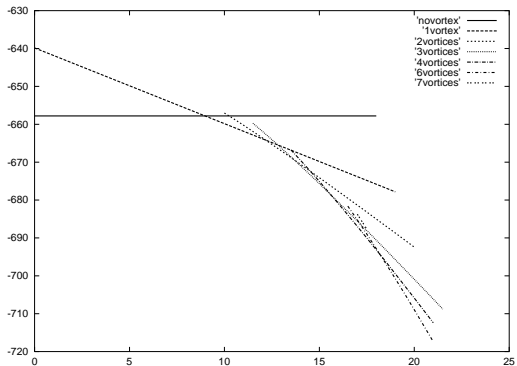}}
\caption{The energy vs. $\Omega$ curves for $\lambda=1.0$}
\label{figener1}
\end{figure}

\begin{figure}
\vspace{0.1in}
\centerline{\epsfxsize=2.8in\epsfbox{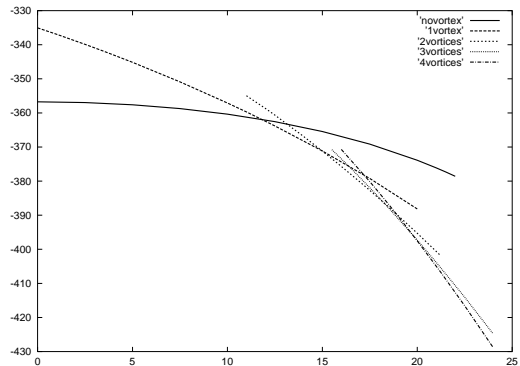}}
\caption{The energy vs. $\Omega$ curves  for $\lambda=1.5$}
\label{figener2}
\end{figure}

We now discuss the energy diagrams in relation to the discussion
of the critical velocities given in the earlier sections. In Figures
\ref{figener1} and \ref{figener2}, we have plotted the energy given by
(\ref{be2d}) as a function of $\O$ for the various branches of solutions
(according to the number of vortices). Again, $\ep=0.02$ in our computation.

\paragraph{The case of the disc (Figure \ref{figener1})}

As discussed earlier,  the vortex-less solution, in the case of the
disc, exists for all values of $\O$ and is independent of $\O$, we
thus have a constant line for its energy. These vortex-free solutions
are the global energy minimizers for small $\O$ ($\O<9.3$) whereas
for $\O>9.3$, they have larger energy than the one-vortex solution. 
 For multi-vortex solutions, we see that each becomes the global 
energy minimizer for a range of values of $\O$.
 It is interesting to compare this result with the value found in
section 2 where $\O_1=9.8$. Similarly, we obtain from (\ref{On}) that
$\O_n-\O_{n-1}\simeq 1.77$. Though the value of $\O_1$ is slightly
overestimated, the difference $\O_n-\O_{n-1}$ looks good for small $n$:
the numerics indicate $\O_2=12.0$, $\O_3=13.6$, $\O_4=15.8$ and our
theoretical computations yield $\O_2=11.6$, $\O_3=13.4$, $\O_3=15.2$.

However, when $\O$ is increased from 0, we saw that we stay on the
vortex free branch up to $\O=19$. This means that the vortex free
solution is a local minimizer up to this value. We do not have any
theoretical estimate for this value of local minimum. 
 Similarly, for multi-vortex solutions,  hysteresis
loops are present.
For the solution with six vortices, there are two possible
configurations, one with all six on a concentric circle, one
with only five on a concentric circle while the remaining vortex
at the center of the disc. This occurs for $\O=22.5$.
Though the difference  in the energy is hardly noticeable, the solution
with a center vortex does have a smaller energy value.

\paragraph{The case of an ellipse (Figure \ref{figener2}) }
Now, we discuss the elliptical case with $\la=1.5$. The energy versus
$\O$ curves are given in Figure \ref{figener2}.

The vortex-less solution, in the case of the
ellipse, is no longer independent of $\O$ as in the case of the
disc, as illustrated by the dependence its energy on $\O$. 
Aside from that, the hysteresis phenomena also occurs just like for the 
case of a disc.

Our numerics indicate that $\O_1=12 $, $\O_2=15$.
It is interesting to compare this result with the value found in section
2: we obtain that $\O_1\simeq 13$ from (\ref{On}) that
$\O_n-\O_{n-1}\simeq 2.88$. Though the value of $\O_1$ is slightly
overestimated, the difference $\O_n-\O_{n-1}$ looks good again since we
find $\O_2=14.9$.

\subsection{Displacement of the vortex from the center}
Based on the earlier estimate (\ref{deltaEellipse}) 
for the one-vortex solution, we see that
for small $\O$, 
the displacement of the vortex away from the center leads to the drop of
energy ($\O <4.9$). For slightly bigger $\O$ a vortex at the center
is a local minimum ($\O <9.8$). Let us analyze the time dependent
equation using a vortex off center as initial condition and let us
examine how it evolves. We find that in the case 
of the disk, for $\O< 7$, a displacement on the size of a tenth of the
radius causes the displaced center vortex in the one-vortex solution
to move towards the boundary. For $\O>10$, 
the center vortex in the one-vortex solution
moves back towards the center if under a displacement on the size of a 
third of the radius. For intermediate values of $\O$, the vortex moves
back to the center for small displacement but moves towards boundary
for large displacement.

\begin{figure}[htb]
\centerline{\epsfxsize=2.5in\epsfbox{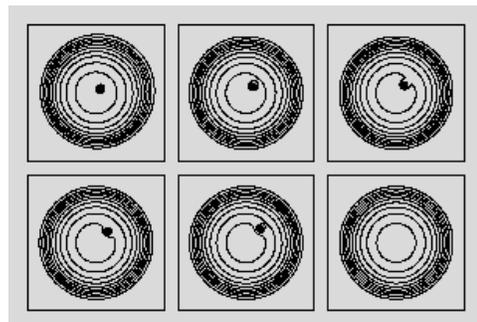}}
\caption{Perturbing vortex away from center for $\Omega=0$}\label{figp1}
\end{figure}

\begin{figure}[htb]
\centerline{\epsfxsize=2.8in\epsfbox{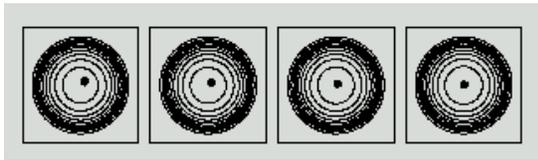}}
\vspace{0.in}
\caption{Perturbing vortex away from center for $\Omega=7.9$}\label{figp2}
\end{figure}

The following two pictures (Figures \ref{figp1} and \ref{figp2}) show,
in the case of a disc and a small displacement, 
the marching away of the
center vortex for the solution with $\O=0$ (starting from top row then
to the bottom row) and the marching back to the
center for the solution with $\O=7.9$. 

\noindent

\section{Conclusion}

We have presented a new framework for the study of the Gross Pitaevskii
energy  in the Thomas Fermi limit: we have defined a nondimensionnalized parameter $\ep$ and we have computed theoretically an asymptotic
development of the energy, the critical thermodynamic velocities of
nucleation of vortices and the location of vortices. This  extends the
results of \cite{CD}. We have proposed and implemented time
integration schemes which enjoy the norm-preserving and
energy-descreasing features and thus useful for studying the stability and
metastability of solutions.
We have also presented
energy diagrams computed numerically for the various vortex
solutions. We have noticed that our theoretical predictions for critical
thermodynamic velocities are quite consistent with the numerics which
encourages us to think that our approximation $\ep$ small is correct
though $\lep$ is not small. 
 
In our computation, we have
taken $\ep=0.02$ and the number of vortices we have observed ranges from
0 to 10 for $\O$ between $0$ and $25$.

Our aim for a future work is to use the results of \cite{PR} for
Ginzburg-Landau in dimension 3 to get results for rotating Bose Einstein
condensates in dimension 3, and especially the critical velocities and
an asymptotic development of the energy. In particular, it is obtained
in \cite{SF} that $\O_s=3/5 \O_1$ for a vortex line parallel to the $z$
axis, but we would like to get an expression of the energy for any type
of vortex line. We will also carry out further numerical simulations
in 3 dimension to compare with experimental data.

\begin{acknowledgments}
\vspace{-0.2in}
The authors would like to thank Y.Castin for very
interesting discussions and also T.Riviere, E.Sandier, S.Serfaty. This
work is supported in part by a joint France-Hong Kong research grant.
\end{acknowledgments}

\end{document}